\documentclass[showpacs,prl,onecolumn,aps,superscriptaddress,preprintnumbers,nofootinbib]{revtex4}
\usepackage[T1]{fontenc}
\usepackage[latin9]{inputenc}
\usepackage{amsmath,amssymb}
\usepackage{epsfig}
\usepackage{bm} 
\usepackage{graphicx}
\usepackage{mathrsfs}
\usepackage{amsmath}
\usepackage{amsfonts}
\usepackage{epstopdf}
\usepackage{color}
\usepackage{pifont}
\usepackage{relsize}
\usepackage{mathtools}
\def\slashchar#1{\setbox0=\hbox{$#1$}     		
   \dimen0=\wd0                                 	
   \setbox1=\hbox{/} \dimen1=\wd1               	
   \ifdim\dimen0>\dimen1                        	
      \rlap{\hbox to \dimen0{\hfil/\hfil}}      	
      #1                                        	
   \else                                        	
      \rlap{\hbox to \dimen1{\hfil$#1$\hfil}}   	
      /                                         	
   \fi}

\renewcommand{\vec}{\boldsymbol}
\newcommand{\beq}{\begin{equation}}
\newcommand{\eeq}{\end{equation}}
\newcommand{\bea}{\begin{eqnarray}}
\newcommand{\eea}{\end{eqnarray}}
\newcommand{\ba}{\begin{array}}
\newcommand{\ea}{\end{array}}

\def\eq#1{{Eq.~(\ref{#1})}}
\def\fig#1{{Fig.~\ref{#1}}}
\newcommand{\bas}{\bar{\alpha}_S}

\newcommand{\nn}{\nonumber}
\newcommand{\bg}{ \bar{\gamma}}

\newcommand{\Lb}{\left(}
\newcommand{\Rb}{\right)}
\newcommand{\h}{\frac{1}{2}}

\newcommand{\pom}{I\!\!P}

\newcommand{\intl}{\int\limits}
\begin{document}

\title{ Summing large Pomeron loops in the saturation region:  amplitude and  multiplicity   distributions  for dipole-dipole  scattering }

\author{Eugene Levin}
\email{leving@tauex.tau.ac.il}
\affiliation{Department of Particle Physics, Tel Aviv University, Tel Aviv 69978, Israel}

\date{\today}

\pacs{13.60.Hb, 12.38.Cy}

\begin{abstract}
In this paper we found  the parton densities at high energies. Their expressions  stems from our attempts to reconcile the exact solution to the Balitsky-Kovchegov (BK) equation, which describes the rare fluctuation in the dipole-target scattering, with the fact that this equation sums the 'fan' Pomeron diagrams. Using these densities we can calculate the contributions of large Pomeron loops to dipole-dipole scattering at high energies. We detected that the scattering matrix for this process has the same suppression 
as was predicted by Iancu and Mueller from their estimates of the 'rare' fluctuation in QCD. 
 Applying the Abramovsky,Gribov and Kancheli cutting rules we found  that  the produced gluons are distributed accordingly the KNO (Koba, Nielsen and Olesen) law which leads to the entropy $S_E = \ln(x G(x,Q^2))$. $xG$ has to be in the saturation region at high energies.

 \end{abstract}
\maketitle

\vspace{-0.5cm}
\tableofcontents

\section{Introduction}
 In this paper we sum the large BFKL Pomeron loops\cite{BFKL}\footnote{BFKL stands for Balitsky, Fadin,Kuraev and Lipatov.} for dipole-dipole scattering at high energies. Summing Pomeron loops  has been one of the  difficult problems in the Color Glass Condensate (CGC) approach, without solving this  we cannot consider the dilute-dilute and dense-dense parton densities collisions. As recently shown\cite{KLLL1,KLLL2}, even the Balitsky-Kovchegov (BK) equation, that governs the dilute-dense parton density  scattering (deep inelastic scattering (DIS) of electron with proton), has to be modified due to  contributions of  Pomeron loops.  However, in spite of intensive work 
\cite{BFKL,KOLEB,MUSA,LETU,LELU,LIP,KO1,LE11,RS,KLremark2,SHXI,KOLEV,nestor,LEPRI,LMM,LEM,MUT,MUPE,IIML,LIREV,LIFT,GLR,GLR1,MUQI,MUDI,Salam,NAPE,BART,BKP,MV, KOLE,BRN,BRAUN,B,K,KOLU,JIMWLK1,JIMWLK2,JIMWLK3, JIMWLK4,JIMWLK5,JIMWLK6,JIMWLK7,JIMWLK8,AKLL,KOLU11,KOLUD,BA05,SMITH,KLW,KLLL1,KLLL2,kl,LEPR,LE1,LE2}, this problem  has  not been solved.  

\begin{figure}[ht]
    \centering
  \leavevmode
      \includegraphics[width=7cm]{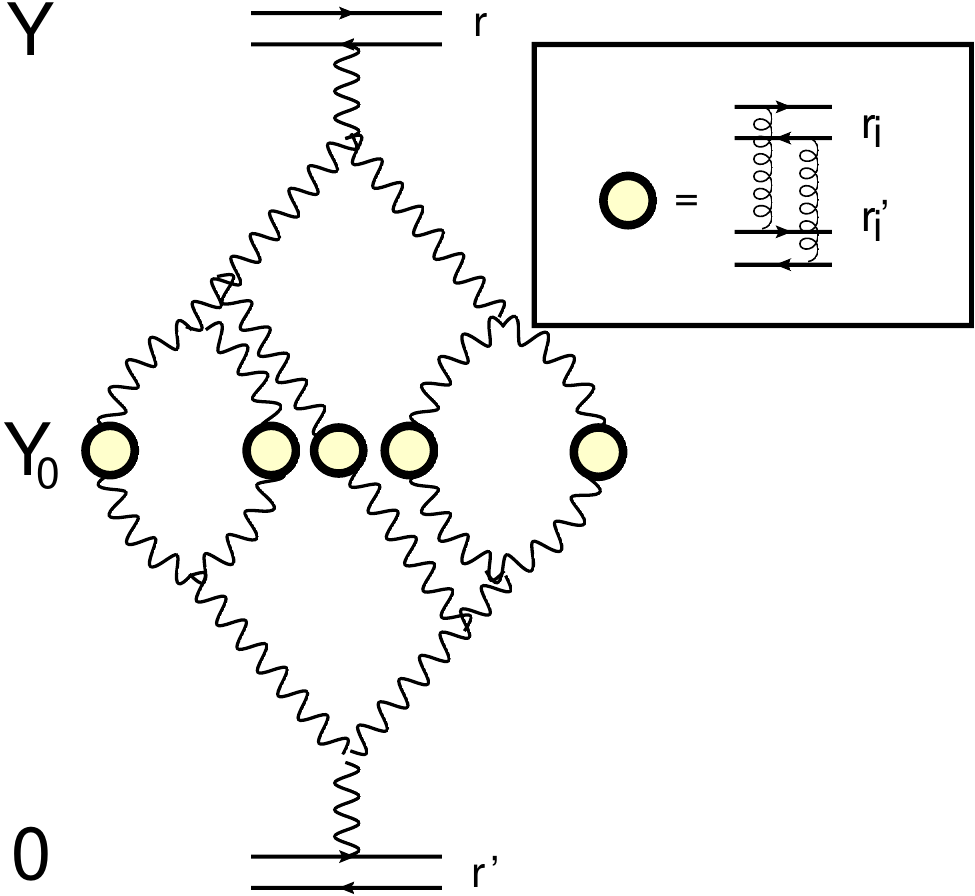}  
      \caption{ Summing  large Pomeron loops. The wavy lines denote the  BFKL Pomeron exchanges.  The circless  denote the amplitude $\gamma$.  }
\label{mpsi}
   \end{figure}

 We sum the large Pomeron loops using the $t$-channel unitarity, which has been rewritten in the convenient form for the dipole approach to CGC in Refs.\cite{MUSA,Salam,IAMU,IAMU1,KOLEB,MUDI,LELU,KO1,LE11}(see \fig{mpsi}).
 The analytic expression takes the form
       \cite{LELU,KO1,LE1}:  \bea \label{MPSI}
     && A\Lb Y, r,R ;  \vec{b}\Rb\,=\\
     &&\,\sum^\infty_{n=1}\,\Lb -1\Rb^{n+1}\,n!\int  \prod^n_i d^2 r_i\,d^2\,r'_i\,d^2 b'_i 
     \int \!\!d^2 \delta b_i\, \gamma^{BA}\Lb r_1,r'_i, \vec{b}_i -  \vec{b'_i}\equiv \delta \vec{b} _i\Rb 
    \,\,\rho_n\Lb Y - Y_0, \{ \vec{r}_i,\vec{b}_i\}\Rb\,\rho_n\Lb Y_0, \{ \vec{r}'_i,\vec{b}'_i\}\Rb \nn
      \eea
  $\gamma^{BA}$ is the scattering amplitude of two dipoles in the Born approximation of perturbative QCD.  The parton densities $\rho_i Y , \{ \vec{r}_i,\vec{b}_i\}$ have been introduced in Ref.\cite{LELU}  as follows:
\beq \label{PD}
\rho_n(r_1, b_1\,
\ldots\,,r_n, b_n; Y\,-\,Y_0)\,=\,\frac{1}{n!}\,\prod^n_{i =1}
\,\frac{\delta}{\delta
u_i } \,Z\left(Y\,-\,Y_0;\,[u] \right)|_{u=1}
\eeq
  where  the generating functional $Z$ is
  \beq \label{Z}
Z\Lb Y, \vec{r},\vec{b}; [u_i]\Rb\,\,=\,\,\sum^{\infty}_{n=1}\int P_n\Lb Y,\vec{r},\vec{b};\{\vec{r}_i\,\vec{b}_i\}\Rb \prod^{n}_{i=1} u\Lb \vec{r}_i\,\vec{b}_i\Rb\,d^2 r_i\,d^2 b_i
\eeq
 where $u\Lb \vec{r}_i\,\vec{b}_i\Rb \equiv\,u_i$ is an arbitrary function and $P_n$ is the probability to have $n$ dipoles with the  given kinematics.
 The initial and  boundary conditions for the BFKL cascade which stems from one dipole has 
the following form for the functional $Z$:
\begin{subequations}
\bea
Z\Lb Y=0, \vec{r},\vec{b}; [u_i]\Rb &\,\,=\,\,&u\Lb \vec{r},\vec{b}\Rb;\label{ZIC}\\
Z\Lb Y, r,[u_i=1]\Rb &=& 1; \label{ZSR}
\eea
\end{subequations}
 In \eq{MPSI} $\vec{b}_i\,\,=\,\,\vec{b} \,-\,\vec{b'}_i$.

  For the parton densities, the evolution equations , and  the recurrence relations have been written in Refs.\cite{LELU,LE1} and the attempts to solve them have been made in Refs.\cite{LE1,LE2}. In this paper we suggest a different way to find these densities. The BFKL  parton cascade leads to the non-linear Balitsky-Kovchegov (BK) equation \cite{B,K}. In Ref.\cite{LETU} the analytical solution to this equation in the saturation region was found:
 \beq\label{I1}
 N^{\rm DIS} \Lb z= r^2 Q^2_s\Lb Y, b\Rb\Rb\,\,=\,\,1\,\,-\,\,C(z)\exp\Lb - \frac{z^2}{2\,\kappa}\Rb
 \eeq 
 where $z = \ln\Lb r^2 Q^2_s(Y)\Rb$\footnote{We will discuss the definition of $z$ in more detail  below in \eq{zz}.} One can see that  this solution
  shows the geometric scaling behaviour \cite{GS}  being a function of one variable. Function $C\Lb z\Rb$ is a smooth function  which can be considered as a constant in our approach.  As one can see from \eq{I1}   it turns out that 
 BK equation  leads to a new dimensional scale: saturation momentum\cite{GLR}  which has the following $Y$ dependence\cite{GLR,MUT,MUPE}:
 \beq \label{QS}
 Q^2_s\Lb Y, b\Rb\,\,=\,\,Q^2_s\Lb Y=0, b\Rb \,e^{\bas\,\kappa \,Y,-\,\,\frac{3}{2\,\gamma_{cr}} \ln Y }
 \eeq 
 where $Y=0$ is the initial value of rapidity and $\kappa$ and $\gamma_{cr}$   are determined by the following equations\footnote{$\chi\Lb \gamma\Rb$ is the BFKL kernel\cite{BFKL} in anomalous dimension ($\gamma$) representation.$\psi$ is the Euler psi -function (see Ref.\cite{RY} formula {\bf 8.36}). }:
  \beq \label{GACR}
\kappa \,\,\equiv\,\, \frac{\chi\Lb \gamma_{cr}\Rb}{1 - \gamma_{cr}}\,\,=\,\, - \frac{d \chi\Lb \gamma_{cr}\Rb}{d \gamma_{cr}}~
\eeq
where $\chi\Lb \gamma\Rb$ is given by
\beq \label{CHI}
\omega\Lb \bas, \gamma\Rb\,\,=\,\,\bas\,\chi\Lb \gamma \Rb\,\,\,=\,\,\,\bas \Lb 2 \psi\Lb 1\Rb \,-\,\psi\Lb \gamma\Rb\,-\,\psi\Lb 1 - \gamma\Rb\Rb\eeq 
     \begin{figure}[ht]
    \centering
  \leavevmode
      \includegraphics[width=7cm]{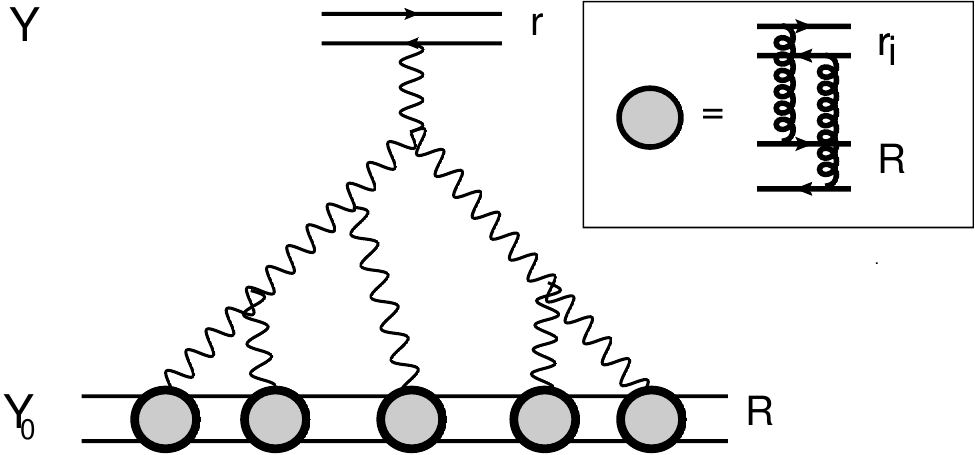}  
      \caption{The non-linear BK equation. The wavy lines denote the  BFKL Pomeron exchanges.  The circles  denote the amplitude $\gamma$ for interaction of the dipole with the target.  }
\label{mpsi1}
   \end{figure}

In the next section we show how to reconcile this solution with the fact that the BK equation is summing the 'fan' diagrams of \fig{mpsi1} in the BFKL Pomeron calculus. We present \eq{I1} as a sum of many Pomerons exchanges and in doing so we find the parton densities $\rho_n$.

  In section III we estimate the scattering amplitude using \eq{MPSI}  and the parton densities that we found in section II . Sections IV,V and VI are devoted to finding the multiplicity of the produced gluons. 
  We approach this problem using the  fact that the BFKL Pomeron generates the cross sections of the produced gluon that are  distributed accordingly the Poisson law.  Based on the AGK\footnote{AGK stands for Abramovsky, Gribov and Kancheli}\cite{AGK} cutting rules we found  the  multiplicity distributions for both interactions. We showed that  in  deep inelastic scattering(DIS) and in dipole-dipole interactions at high energies the produced gluons are  distributed in accord with the  KNO\footnote{KNO stands for  Z. Koba, H.B. Nielsen and P. Olesen}\cite{KNO}  law  with the same average number of dipoles. This number is determined by the gluon structure function in the saturation region.  We found  that the entropy of produced gluons turns out to be the same for both reactions at high energies and it is equal to $S_E=\ln\Lb x G(x,Q^2)\Rb$ in the agreement of the results of Ref.\cite{KHLE}.
 In conclusion we summarize our results and discuss possible flaw in our approach,

    \begin{boldmath}
    \section{Dipole densities $\rho_n\Lb r,b, \{r_i,b_i\}\Rb$ } 
    \end{boldmath}
    
As has been mentioned the BK nonlinear equation is the sum of 'fan' diagrams of the BFKL Pomeron calculus\cite{KOLEB,GLR,BRN,BRAUN} (see \fig{mpsi1}). At high energies (deeply in the saturation region) the exchange of one BFKL Pomeron can written in the following way\cite{GLR,MUT}:
\beq \label{DD1}
G_{\pom}\Lb z \Rb  = N_0 e^{\bg \,z} ;~~~~~~~~G_{n \pom} = \Lb G_{\pom}\Lb z \Rb \Rb^n\eeq
where $\bg = 1 - \gamma_{cr}$ , $N_0$ is a constant and $G_{n \pom}$ is the Green's function for the exchanges of $n$ Pomerons.  $z$ in \eq{DD1} is defined as (see \fig{mpsi1})
\beq \label{zz}
z\,\,=\,\,\bas \frac{\chi\Lb \bg\Rb}{\bg} \Lb Y \,-\,Y_0\Rb \,\,+\,\,\xi_{r,R}
\eeq 
where
$\xi_{r,R}$ we can find from  the eigenfunction of the BFKL equation with the eigenvalue $\bas \chi\Lb \gamma\Rb$
 (the scattering amplitude of two dipoles with sizes $r$ and $R$) which  has the
 following form \cite{LIP}
\beq \label{XI}
\phi_\gamma\Lb \vec{r} , \vec{R}, \vec{b}\Rb\,\,\,=\,\,\,\Lb \frac{
 r^2\,R^2}{\Lb \vec{b}  + \h(\vec{r} - \vec{R})\Rb^2\,\Lb \vec{b} 
 -  \h(\vec{r} - \vec{R})\Rb^2}\Rb^\gamma =e^{\gamma\,\xi_{r,R} }~~\mbox{with}\,\,0 \,<\,Re\gamma\,<\,1
 \eeq

Comparing \eq{I1} with \eq{DD1} one can see that we have to solve the first problem: how to rewrite \eq{I1} as the sum of multi Pomeron exchanges.
We suggest the following formula  for S-matrix ($ S = 1 - N$):
\begin{subequations}
\bea
S\Lb z\Rb\,\,&=&\,\,C\intl^{\epsilon + i \infty}_{\epsilon - i \infty} \frac{d \omega}{2\,\pi\,i} e^{ \frac{ 
\bg^2\,\kappa}{2} \omega^2} \Lb 1 + \frac{1}{N_0} G_{\pom}\Lb z \Rb\Rb^{-\omega} \,\,=\,\,\frac{C}{\sqrt{2\,\pi\,\bg^2\,\kappa}}\exp\Lb-\,\ln^2\Lb 1 + \frac{G_{\pom}\Lb z\Rb}{N_0} \Rb\Rb \label{DD2}\\
&\xrightarrow{z \gg 1}& \frac{C}{\sqrt{2\,\pi\,\bg^2\,\kappa}}e^{ - \frac{z^2}{ 2\, \kappa}} \,\,
=\,\, C \,\intl^{\epsilon + i \infty}_{\epsilon - i \infty} \frac{d \omega}{2\,\pi\,i} e^{ \frac{ 
\bg^2\,\kappa}{2}\,\omega^2}   \sum^\infty_{n = 0} \frac{(-1)^n}{n!} \frac{\Gamma\Lb \omega +n\Rb}{\Gamma\Lb \omega \Rb}\Lb \frac{G_{\pom}\Lb z\Rb}{N_0}\Rb^n\label{DD3} \eea
\end{subequations}

\eq{DD3} gives the solution of \eq{I1} as the sum over multi Pomeron exchanges. Since $G_{n \pom}$ increases as $e^{n\, \bg \kappa Y}$ \eq{DD3} is the asymptotic series which can be summed using the Borel 
resummation \cite{BOREL}.  Actually, \eq{DD2} and \eq{DD3}  show how to sum this asymptotic series.

On the other hand $S\Lb z\Rb$ can be written as \cite{KOLEB,K,LELU}
\beq \label{DD4}
S\Lb z\Rb\,\,=\sum^{\infty}_{n=0} (-1)^n \int \prod^n_{i=1} \gamma^{BA}\Lb r_i,R; b_i\Rb\,d^2r_i\,d^2 b_i\,\,\rho_n\Lb r,b, \{ r_i,b_i\}\Rb 
\eeq
where $\gamma^{BA}$ is the amplitude of interaction of the dipole with rapidity $Y_0$ and the size $r_i$ with the target  of size $R$ (see \fig{mpsi1}) in the Born approximation of perturbative QCD.

Comparing   \eq{DD4} with \eq{DD3} one can see that
\beq \label{DD5}
\rho_n\Lb Y - Y_0;  r,b, \{ r_i,b_i\}\Rb = C \,\intl^{\epsilon + i \infty}_{\epsilon - i \infty} \frac{d \omega}{2\,\pi\,i} e^{ \frac{ 
\bg\,\kappa}{2}\,\omega^2}\frac{(-1)^n}{n!} \frac{\Gamma\Lb \omega +n\Rb}{\Gamma\Lb \omega \Rb}  \prod^n_{i=1} \frac{G_{\pom}\Lb z_i\Rb}{N_0}
\eeq
where
\beq \label{zi}
z_i \,\,=\,\,\,\,\bas \frac{\chi\Lb \bg\Rb}{\bg} \Lb Y \,-\,Y_0\Rb \,\,+\,\,\xi_{r,r_i}
\eeq
 with $\xi_{r, r_i}$ from \eq{XI} in which $R $ is replaced by   $r_i$.
 
 It is instructive to note that \eq{DD5} satisfies the evolution equation for $\rho_n$\cite{LELU} at large values of $z$.

  ~
  
  ~

        ~
      
     \begin{boldmath}
     \section{Scattering amplitude}
      \end{boldmath}

      
Using \eq{DD5} for the parton densities we can find the sum of large Pomeron loops using \eq{I1}. It takes the form:
\bea \label{SA1}
&&S\Lb z\Rb =\\
&&C^2\sum^\infty_{n=0} \frac{\Lb - 1\Rb^n}{n!}\!\!\!  \intl^{\epsilon + i \infty}_{\epsilon - i \infty}\!\!\! \frac{ d \omega_1}{2\,\pi\,i} \!\!\!\intl^{\epsilon + i \infty}_{\epsilon _ i \infty}\!\!\!  \frac{d \omega_2}{2\,\pi\,i} e^{ \frac{ \bg^2 \kappa}{2}\Lb \omega^2_1 + \omega^2_2\Rb}\frac{ \Gamma\Lb \omega_1+n\Rb}{\Gamma\Lb \omega_1\Rb}\frac{ \Gamma\Lb \omega_2+n\Rb}{\Gamma\Lb \omega_2\Rb}\intl\!\!\! d^2 r_i d^2 r'_i d^2b_i d^2b'_i\prod^n_{i=1} 
\frac{G_{\pom}\Lb z_i\Rb}{N_0}\gamma^{BA}\Lb r_i,r'_i,\delta b\Rb\frac{G_{\pom}\Lb z'_i\Rb}{N_0}\nn\eea

In   \eq{SA1}  $z_i$ is given in \eq{zi} while $z'_i \,\,=\,\,\,\,\bas \frac{\chi\Lb \bg\Rb}{\bg} \Lb Y_0\Rb \,\,+\,\,\xi_{r',r'_i}$. $\xi_{r',r'_i}$ is equal to $\xi_{r,R}$ of \eq{XI} where $r$  and $R$ are  replaced by $r'$ and $r'_i$, respectively.

\eq{SA1} can be simplified using t-channel unitarity constraints for the BFKL Pomeron\cite{BFKL}.  In the coordinate representation it can be written in two equivalent forms:
\bea \label{SA2}
G_{\pom} \Lb Y, r,R ;  \vec{b}\Rb \,\,&=&\,\,\intl d^2 r_i d^2 b_i d^2r'_i d^2 b'_i 
\rho_1\Lb Y - Y_0; r, b -b_i, r_i\Rb \gamma^{BA}\Lb r_i,r'_i,\delta b\Rb \rho_1\Lb  Y_0; r,' b'_i, r'_i\Rb\nn\\
&\,\,=\,\,&
 \,\,\frac{1}{4\,\pi^2}\int \frac{d^2 r_i}{r^4_i} \,d^2 b_i 
   \,   G_{\pom} \Lb Y - Y_0, r, r_i;  \vec{b}\ - \vec{b}_i \Rb \,\, G_{\pom} \Lb Y_0,  r_i;  \ \vec{b}_i \Rb  
   \eea  
   The first equation is written in Ref.\cite{MUSA}, while the second has been discussed in Ref. \cite{CLMSOFT}.

  Using \eq{SA2} we reduce \eq{SA1} to the following expression:
  \beq \label{SA3}
S\Lb z'\Rb =
C^2\sum^\infty_{n=0} \frac{\Lb - 1\Rb^n}{n!}\!\!\!  \intl^{\epsilon + i \infty}_{\epsilon - i \infty}\!\!\! \frac{ d \omega_1}{2\,\pi\,i} \!\!\!\intl^{\epsilon + i \infty}_{\epsilon _ i \infty}\!\!\!  \frac{d \omega_2}{2\,\pi\,i} e^{ \frac{ \bg^2 \kappa}{2}\Lb \omega^2_1 + \omega^2_2\Rb}\frac{ \Gamma\Lb \omega_1+n\Rb}{\Gamma\Lb \omega_1\Rb}\frac{ \Gamma\Lb \omega_2+n\Rb}{\Gamma\Lb \omega_2\Rb}\Lb \frac{G_{\pom}\Lb z'\Rb}{N_0}\Rb^n
\eeq
 with 
 \beq \label{SA4}
 z' \,\,= \,\,\,\,\bas \frac{\chi\Lb \bg\Rb}{\bg} \,Y \,\,+\,\,\xi_{r,r'}    
 \eeq   
  $\xi_{r,r'}$ is equal to $\xi_{r,R}$ of \eq{XI} where $r$  and $R$ are  replaced by $r$ and $r'$, respectively.
  
  Plugging in \eq{SA4} the integral representation for $\Gamma$-function: $\Gamma\Lb \omega_2\Rb = \int^\infty_0 d t e^{-t} t^{\omega_2 -1}$ we obtain:
    \beq \label{SA5}
S\Lb z'\Rb =C^2
\sum^\infty_{n=0} \frac{\Lb - 1\Rb^n}{n!}\!\!\!  \intl^{\epsilon + i \infty}_{\epsilon - i \infty}\!\!\! \frac{ d \omega_1}{2\,\pi\,i} \!\!\!\intl^{\epsilon + i \infty}_{\epsilon _ i \infty}\!\!\!  \frac{d \omega_2}{2\,\pi\,i} e^{ \frac{ \bg^2 \kappa}{2}\Lb \omega^2_1 + \omega^2_2\Rb}\intl^\infty_0 d t  e^{-t}  t^{\omega_2 -1}  \frac{ \Gamma\Lb \omega_1+n\Rb}{\Gamma\Lb \omega_1\Rb\Gamma\Lb \omega_2\Rb}\Lb t\, \frac{G_{\pom}\Lb z'\Rb}{N^2_0}\Rb^n
\eeq
  After summation over $n$ we have
          \bea \label{SA6}
S\Lb z'\Rb &=&C^2
 \intl^{\epsilon + i \infty}_{\epsilon - i \infty}\!\!\! \frac{ d \omega_1}{2\,\pi\,i} \!\!\!\intl^{\epsilon + i \infty}_{\epsilon _ i \infty}\!\!\!  \frac{d \omega_2}{2\,\pi\,i} e^{ \frac{ \bg^2 \kappa}{2}\Lb \omega^2_1 + \omega^2_2\Rb}\frac{ 1}{\Gamma\Lb \omega_2\Rb}\intl^\infty_0 d t  e^{-t} \Lb 1\,+\,t  \frac{G_{\pom}\Lb z'\Rb}{N^2_0} \Rb^{- \omega_1 }\,t^{\omega_2 -1}\nn\\
 &=&\!\!C^2 \!\! \!\!\!\!\intl^{\epsilon + i \infty}_{\epsilon - i \infty}\!\!\! \frac{ d \omega_1}{2\,\pi\,i} \!\!\!\intl^{\epsilon + i \infty}_{\epsilon _ i \infty}\!\!\!  \frac{d \omega_2}{2\,\pi\,i} e^{ \frac{ \bg^2 \kappa}{2}\Lb \omega^2_1 + \omega^2_2\Rb}\Bigg\{\frac{\Gamma (\text{$\omega $2}-\text{$\omega $1})}{N_0 \Gamma\Lb \omega_2\Rb}e^{-\bg\omega_1\,z'} \, _1F_1\left(\text{$\omega $1};\text{$\omega $1}-\text{$\omega $2}+1;e^{-\bg z'} N_0\right)\,\, +\,\,\omega_1\longleftrightarrow  \omega_2\Bigg\} 
 \eea
 We introduce new variables $\h( \omega_1 + \omega_2) = \Sigma$ and $\h( \omega_1 - \omega_2) = \Delta$ for investigating the limit of large $z'$. \eq{SA6} takes the form:
 \beq \label{SA7} 
S\Lb z'\Rb 
\xrightarrow{z' \gg 1} C^2\intl^{\epsilon + i \infty}_{\epsilon - i \infty}\!\!\! \frac{ d \Delta}{2\,\pi\,i} \!\!\!\intl^{\epsilon + i \infty}_{\epsilon _ i \infty}\!\!\!  \frac{d \Sigma}{2\,\pi\,i} e^{ \bg^2 \kappa\Lb \Sigma^2 +\Delta^2\Rb}e^{- \bg\,\Sigma\,z'}\Bigg\{\frac{ \Gamma \Lb -\Delta\Rb}{N_0 \Gamma\Lb \Sigma - \Delta\Rb}e^{-\bg\Delta\,z'}\,\,  +\,\,\Delta\leftrightarrow -\Delta\Bigg\}
 \eeq
 Integral over $\Sigma$ leads to $\Sigma \sim z$ while integral over $\Delta $ leads to $\Delta \sim {\rm Const}$. Indeed, for small $\Delta$ the expression in $\{ \dots \}$  looks as follow for imaginary $\Delta = {\rm  i \delta}$  with $\delta \ll z$:
  \beq \label{SA71} 
  \{ \dots \}  \,=\,\frac{1}{\Gamma\Lb \Sigma\Rb\,N_0} \frac{ \sin\Lb \bg \delta \,z'\Rb}{\delta}
  \eeq
  This function has a  maximum at $\delta =0$ where it is equal to $\bg z'$ and the typical width of the order of $1/(\bg z')$. Therefore, we can replace it by the delta function and reduce \eq{SA7} to the following expression:
  
  \beq \label{SA72} 
S\Lb z'\Rb 
\xrightarrow{z' \gg 1} C^2 \!\!\!\intl^{\epsilon + i \infty}_{\epsilon _ i \infty}\!\!\!  \frac{d \Sigma}{2\,\pi\,i} e^{  \bg^2 \kappa\, \Sigma^2 }e^{-\bg\,\Sigma\,z'}\frac{ 1}{N_0 \Gamma\Lb \Sigma \Rb}  \eeq

  The integrals over $\Sigma$  can be taken by the method of  steepest descent  since this approximation  leads to large saddle points: $\Sigma \sim z'$ .
  
  The resulting $S\Lb z'\Rb $ takes the form:
  \beq \label{SA8}
  S\Lb z'\Rb =C'^2 \frac{1}{\Gamma\Lb \frac{z'}{\bg \,\kappa}\Rb} \exp\Lb - \frac{ z'^2}{4 \kappa}\Rb
  \eeq
  where all constants as well as smooth functions of $z'$ are  included in $C'$.
  
  Actually, we cannot   guarantee  term  $ \frac{1}{\Gamma\Lb \frac{z'}{\bg \,\kappa}\Rb}$  in  this equation, since the solution of \eq{I1} absorbs such kind of correction in smooth function $C(z)$. Hence,  we can conclude that the sum of the large Pomeron loops leads to the  S-matrix in perfect agreement with the estimates of 'rare' fluctuation, given in Ref.\cite{IAMU}.
         ~

         ~

         ~
     \begin{boldmath}
     \section{Multiplicity distribution  of produced gluons for BK scattering amplitude }
      \end{boldmath}

  
         ~
      In this section we are going to use \eq{DD3} to obtain the multiplicity distribution of produced gluons for BK scattering amplitude.    
Our basic ideas are  outlined  in Ref.\cite{MUSA} and our approach consists of two steps. First, we recall that it is proven in Refs.\cite{BFKL} that the $s$-channel unitarity for the BFKL Pomeron has the form:
  \beq \label{MD1}
2\, { \rm Im} \,G_{\pom}\Lb z \Rb\,\,=\,\,\sigma^{\mbox{\tiny BFKL}}_{in}(z)
\eeq
 where $G_{\pom}$ is the Green's function for the BFKL Pomeron (see \eq{DD1}) and  $\sigma^{\mbox{\tiny BFKL}}_{in}(z)$ is the inelastic cross sections of  produced gluons with mean multiplicity $\bar{n} = \bg \,z$ (see Appendix A) , where $\bg = 1 - \gamma_{cr}$ (see \eq{CHI}). We also know that produced gluons have the Poisson distribution  with this mean multiplicity\cite{LEHP}( (see Appendix A) .
 
 The second step is the AGK cutting rules\cite{AGK}, which allow us to calculate the imaginary part of   the scattering amplitude, that determines the cross sections, through the powers of ${ \rm Im} \,G^{\mbox{\tiny BFKL}}\Lb z\Rb$\footnote{In the widespread slang this contribution is called by   cut Pomeron.}. Our master formula takes the form of convolution for the cross section of produced $n$  gluons:
 \beq \label{MD2}
 \sigma_n\Lb z \Rb\,\,=\,\,\sum_k\underbrace{ \sigma_k^{AGK}\Lb z \Rb}_{ \propto\,\Lb{\rm Im} G_{\pom} \Rb^k}\underbrace{ \frac{\Lb k \, \bg \,z\Rb^n}{n!} e^{ -k\, \bg\,z}}_{\mbox{Poisson distribution}}
\eeq

The AGK cutting rules \cite{AGK} allows us to calculate the contributions of $n$-cut Pomerons if we know $F_k$: the contribution of the exchange of $k$-Pomerons to  the cross section. They take the form:
   \begin{subequations} 
    \bea 
n\,\geq\,1:\sigma^k_n\Lb Y, \xi_{r,R}\Rb&=& (-1)^{n -k}\frac{k!}{(n - k)!\,n!}\,2^{k}\, F_k(Y,\xi_{r,R})\label{AGKK}\\
n\,=\,0:\sigma^k_0\Lb Y\Rb&=&\Lb -1\Rb^k \Bigg(2^k\,\,-\,\,2\Bigg) F_k(Y,\xi_{r,R});\label{AGK0}\\
\sigma_{tot}&=&\,\,2 \sum_{k=1}^\infty (-1)^{k+1} \,F_k(Y,\xi_{r,R});\label{XS}\,
\eea
 \end{subequations} 
$\sigma_{tot} = 2\, {\rm Im} A\Lb z\Rb$ where $A$ is the scattering amplitude. $\sigma_0$ is the cross section of diffractive  production of small numbers of gluons which is much smaller than  $\Delta\, Y$.

Plugging $F_k(Y,\xi_{r,R})$ from \eq{DD3} into \eq{AGKK} and using $\Gamma(x) =\int^\infty_0 d t e^{-t} t^{x-1}$ we obtain:
\begin{subequations}\bea 
\sigma_n\Lb z\Rb\,&=&\,\sum_{k=n}^\infty\sigma^k_n\Lb z\Rb\,=\,
 C \,\!\!\!\!\!\!\intl^{\epsilon + i \infty}_{\epsilon - i \infty} \!\!\!\!\!\frac{d \omega}{2\,\pi\,i} e^{ \frac{ 
\bg^2\,\kappa}{2}\,\omega^2}\frac{1}{\Gamma\Lb \omega \Rb} \int^\infty_0\!\!\!\!\!\!d  t  \,t^{\omega - 1}\,e^{-t} \sum^\infty_{k=n } \frac{(-1)^{n-k}}{k!} \Lb \frac{ k!}{(n-k)! \,n!} \Rb\Lb 2\, t\, \frac{G_{\pom}\Lb z\Rb}{N_0}\Rb^k   \label{MD3}\\
&=& C \,\!\!\intl^{\epsilon + i \infty}_{\epsilon - i \infty} \frac{d \omega}{2\,\pi\,i} e^{ \frac{ 
\bg^2\,\kappa}{2}\,\omega^2}\frac{1}{\Gamma\Lb \omega \Rb}\intl^\infty_0\!\! d t\,  e^{-t} t^{\omega - 1} \frac{1}{n!}\Lb 2\, t\, \frac{G_{\pom}\Lb z\Rb}{N_0}\Rb^{n} \exp\Lb -2\, t\, \frac{G_{\pom}\Lb z\Rb}{N_0}\Rb \label{MD4}\\
&=&\frac{C}{n!}\intl^{\epsilon + i \infty}_{\epsilon - i \infty} \frac{d \omega}{2\,\pi\,i} e^{ \frac{ \bg^2\,\kappa}{2}\,\omega^2}\frac{1}{\Gamma\Lb \omega \Rb} \Lb 2\,  \frac{G_{\pom}\Lb z\Rb}{N_0}\Rb^n \Gamma (n+\omega ) \, \Lb 1 + 2  \frac{G_{\pom}\Lb z\Rb}{N_0}\Rb^{- n - \omega} \label{MD5} \\
&\xrightarrow{z \gg 1} &\frac{C}{n!}
\exp\Lb - \frac{n}{2\, \frac{G_{\pom}\Lb z\Rb}{N_0}}\Rb\,\intl^{\epsilon + i \infty}_{\epsilon - i \infty} \frac{d \omega}{2\,\pi\,i} e^{ \frac{ \bg^2\,\kappa}{2}\,\omega^2} \frac{1}{\Gamma\Lb \omega\Rb} \Gamma\Lb n + \omega\Rb \Lb 2\, \frac{G_{\pom}\Lb z\Rb}{N_0}\Rb^{-\omega}\label{MD6}
\eea
\end{subequations}

From \eq{MD6} we see that the typical $n$ are large and of the order of 
$2\, \frac{G_{\pom}\Lb z\Rb}{N_0}$.  Hence we can simplify \eq{MD6}:
\beq \label{MD7}
 \sigma_n = C
\exp\Lb - \frac{n}{2\, \frac{G_{\pom}\Lb z\Rb}{N_0}}\Rb\,\intl^{\epsilon + i \infty}_{\epsilon - i \infty} \frac{d \omega}{2\,\pi\,i} e^{ \frac{ \bg^2\,\kappa}{2}\,\omega^2} \frac{1}{\Gamma\Lb \omega\Rb}  \Lb 2\, \frac{G_{\pom}\Lb z\Rb}{n\,\,N_0}\Rb^{-\omega}\,\,\equiv \,\,\Psi\Lb \frac{n}{ e^{\bg z}}\Rb
\eeq

Therefore, the multiplicity distribution satisfy the KNO scaling behaviour \cite{KNO} with the KNO function that is determined by integral in \eq{MD7}.

For $ \frac{n}{e^{\bg z}}$ either small or large we can take the integral in \eq{MD7} using the method of steepest descent and obtain $\Psi$ in the form

\beq \label{MD7}
\Psi\Lb \frac{n}{ e^{\bg z}}\Rb
 = \frac{C}{\Gamma\Lb\frac{1}{\bg^2 \kappa} \ln\Lb \frac{2 e^{\bg\,z}}{n}\Rb \Rb}\,\exp\Lb - \frac{n}{2\, e^{\bg z}}\Rb\exp\Lb 
 - \frac{1}{2 \,\kappa}  \ln^2\Lb \frac{2 e^{\bg\,z}}{n}\Rb\Rb
 \eeq

The probability to have $n$ cut Pomerons  in the final sate is equal to
\beq \label{MD8}
\mathcal{P}^{AGK}_n = \frac{\sigma^{AGK}_n}{\sum_{n=1}^\infty \sigma^{AGK}_n}=C \frac{1}{ e^{\bg z}}\Psi\Lb \frac{n}{ e^{\bg z}}\Rb
\eeq
We can estimate   the value of constant but we do not do this since we are only able to calculate  the large contributions in the exponent.

For produced gluons \eq{MD2} gives 
\bea \label{MD9} 
\sigma_n &=&\underbrace{\sum_k \frac{C}{\Gamma\Lb\frac{1}{\bg^2 \kappa} \ln\Lb \frac{2 e^{\bg\,z}}{k}\Rb \Rb}\,\exp\Lb - \frac{n}{2\, e^{\bg z}}\Rb\exp\Lb 
 - \frac{1}{2 \,\kappa}  \ln^2\Lb \frac{2 e^{\bg\,z}}{k}\Rb\Rb}_{\sigma^{AGK}_k}\,\underbrace{\frac{\Lb k\,\bg \,z'\Rb}{n!} e^{ - k\,\bg\,z'}}_{Poisson\,\, distribution}\nn\\
 & =& \frac{C}{\Gamma\Lb\frac{1}{\bg^2 \kappa} \ln\Lb \frac{2 e^{\bg\,z}}{k}\Rb \Rb}\,\exp\Lb - \frac{n}{2\, e^{\bg z}}\Rb\exp\Lb 
 - \frac{1}{2 \,\kappa}  \ln^2\Lb \frac{2 e^{\bg\,z}}{k}\Rb\Rb \Big{|}_{k = \frac{n}{ \bg \,z}}  \eea
  In this equation we use that the Poisson distribution is a  rapidly varying   function with typical $n \propto z$ while $\sigma^{AGK}_k$ is almost constant on this scale with characteristic scale  $k \sim 2\exp\Lb \bg z\Rb$.
  
  For the probabilities we have
  \beq \label{MD10}
\mathcal{P}_n = C \frac{\bg \,z}{ e^{\bg z}}\Psi\Lb \frac{\frac{n}{\bg z}}{ e^{\bg z}}\Rb\eeq  
   
           ~

         ~
     \begin{boldmath}
     \section{Multiplicity distribution  of produced gluons for dipole-dipole scattering  }
      \end{boldmath}   
   
      
      Using AGK cutting rules of \eq{AGKK},\eq{AGK0} and \eq{XS} and the  amplitude of dipole-dipole scattering (see \eq{SA5} one can see that $\sigma_n$ is equal to
     
    \bea \label{MDDD1}
\sigma_n\Lb z'\Rb & =&C^2
\sum^\infty_{k=n} \frac{\Lb - 1\Rb^{k -n }}{k!}\Lb \frac{k!}{(k - n)! \,n!}\Rb\!\!\!  \intl^{\epsilon + i \infty}_{\epsilon - i \infty}\!\!\! \frac{ d \omega_1}{2\,\pi\,i} \!\!\!\intl^{\epsilon + i \infty}_{\epsilon _ i \infty}\!\!\!  \frac{d \omega_2}{2\,\pi\,i} e^{ \frac{ \bg^2 \kappa}{2}\Lb \omega^2_1 + \omega^2_2\Rb}\intl^\infty_0 d t e^{-t}  \frac{t^{\omega_1 -1} }{\Gamma\Lb \omega_1\Rb}\,\frac{\Gamma\Lb \omega_2 + k\Rb}{\Gamma\Lb \omega_2\Rb}\Lb 2\,t\, \frac{G_{\pom}\Lb z'\Rb}{N^2_0}\Rb^k\nn\\
&=&  \frac{C^2}{n!}\!\!\!  \intl^{\epsilon + i \infty}_{\epsilon - i \infty}\!\!\! \frac{ d \omega_1}{2\,\pi\,i} \!\!\!\intl^{\epsilon + i \infty}_{\epsilon _ i \infty}\!\!\!  \frac{d \omega_2}{2\,\pi\,i} e^{ \frac{ \bg^2 \kappa}{2}\Lb \omega^2_1 + \omega^2_2\Rb}\intl^\infty_0  d t \, e^{-t}   \frac{t^{\omega_1 -1}}{\Gamma\Lb \omega_1\Rb\,\Gamma\Lb \omega_2\Rb} 
\Gamma\Lb n + \omega_2\Rb
\Lb 2\,t\, \frac{G_{\pom}\Lb z'\Rb}{N^2_0}\Rb^n \Lb  1+ 2\,t\, \frac{G_{\pom}\Lb z'\Rb}{N^2_0}\Rb^{- \omega_2 - n}
\eea
In the region of large $z$ we have
\bea \label{MDDD11}
\sigma_n&= &   \frac{C^2}{n!}\!\!\!  \intl^{\epsilon + i \infty}_{\epsilon - i \infty}\!\!\! \frac{ d \omega_1}{2\,\pi\,i} \!\!\!\intl^{\epsilon + i \infty}_{\epsilon _ i \infty}\!\!\!  \frac{d \omega_2}{2\,\pi\,i} e^{ \frac{ \bg^2 \kappa}{2}\Lb \omega^2_1 + \omega^2_2\Rb}\intl^\infty_0 \!\! d t\, e^{-t}   \frac{t^{\omega_1 -1}}{\Gamma\Lb \omega_1\Rb\,\Gamma\Lb \omega_2\Rb} 
\Gamma\Lb n + \omega_2\Rb
\Lb 2\,t\, \frac{G_{\pom}\Lb z'\Rb}{N^2_0}\Rb^{- \omega_2}\!\!\!\!\!\!\exp\Lb - \frac{n + \omega_2}{2\,t\, \frac{G_{\pom}\Lb z'\Rb}{N^2_0}}\Rb\\
&=& \frac{2 C^2}{n!}\!\!\!  \intl^{\epsilon + i \infty}_{\epsilon - i \infty}\!\!\! \frac{ d \Sigma}{2\,\pi\,i} \!\!\!\intl^{\epsilon + i \infty}_{\epsilon _ i \infty}\!\!\!  \frac{d \Delta}{2\,\pi\,i} e^{ \bg^2 \kappa\Lb \Sigma^2 +\Delta^2\Rb}   \frac{\Gamma\Lb n + \Sigma  -\Delta \Rb}{\Gamma\Lb\Sigma  +\Delta \Rb\,\Gamma\Lb\Sigma  -\Delta  \Rb} 
\Lb 2\, \frac{G_{\pom}\Lb z'\Rb}{N^2_0}\Rb^{- \Sigma}\,\Lb \frac{1}{n +\Sigma - \Delta}\Rb^{- \Delta}  \!\!\!\!\!\! K_{-\Delta}\left(2{\sqrt{\frac{ n  +   \omega_2}{2\, \frac{G_{\pom}\Lb z'\Rb}{N^2_0}}}}\right)\nn
\eea  
From this equation we can see that  $n$ can be large and the typical n$ \sim 
e^{\bg \,z'}$. Bearing this in mind we can reduce  \eq{MDDD11} to the following expression:
\beq \label{MDDD12}
\sigma_n\,=\,C'^2\!\!\!  \intl^{\epsilon + i \infty}_{\epsilon - i \infty}\!\!\! \frac{ d \Sigma}{2\,\pi\,i} \!\!\!\intl^{\epsilon + i \infty}_{\epsilon _ i \infty}\!\!\!  \frac{d \Delta}{2\,\pi\,i}    \frac{e^{  \bg^2 \kappa\Lb \Sigma^2 +\Delta^2\Rb}}{\Gamma\Lb\Sigma  +\Delta \Rb\,\Gamma\Lb\Sigma  -\Delta  \Rb} 
\Lb 2\, \frac{G_{\pom}\Lb z'\Rb}{n\,\,N^2_0}\Rb^{- \Sigma}\,  \!\!\!\!\!\! K_{-\Delta}\left(2{\sqrt{\frac{ n  +   \omega_2}{2\, \frac{G_{\pom}\Lb z'\Rb}{N^2_0}}}}\right) = \Psi\Lb \frac{ n  }{2\, \frac{G_{\pom}\Lb z'\Rb}{N^2_0}}\Rb\eeq
For $\frac{ n}{2\, \frac{G_{\pom}}{N^2_0}} \,\gg\,1$ the KNO function can be calculated since $K_{- \Delta}$ does not depend on $\Delta$. \eq{MDDD12}
takes the form:
\beq \label{MDDD12}
\Psi\Lb \frac{ n  }{2\, \frac{G_{\pom}\Lb z'\Rb}{N^2_0}}\Rb\,=\,C'^2\!\!\!  \intl^{\epsilon + i \infty}_{\epsilon - i \infty}\!\!\! \frac{ d \Sigma}{2\,\pi\,i} \!\!\!\intl^{\epsilon + i \infty}_{\epsilon _ i \infty}\!\!\!  \frac{d \Delta}{2\,\pi\,i}    \frac{e^{ \bg^2 \kappa\Lb \Sigma^2 +\Delta^2\Rb}}{\Gamma\Lb\Sigma  +\Delta \Rb\,\Gamma\Lb\Sigma  -\Delta  \Rb} 
\Lb 2\, \frac{G_{\pom}\Lb z'\Rb}{n\,\,N^2_0}\Rb^{- \Sigma}\,  \!\!\!\!\!\! \exp \left(-2{\sqrt{\frac{ n  }{2\, \frac{G_{\pom}\Lb z'\Rb}{N^2_0}}}}\right)  \eeq

One can see that the typical value of $\Delta \sim Const$ while $\Sigma \sim \ln\left(-2{\sqrt{\frac{ n  }{2\, \frac{G_{\pom}\Lb z'\Rb}{N^2_0}}}}\right) \gg\,\,1$. Note, these estimates confirm our conclusion that $\Sigma \gg \Delta$ in the calculation of the scattering matrix. Taking the integral using the method of steepest descent we obtain:
\beq \label{MDDD13}
\Psi\Lb \frac{ n  }{2\, \frac{G_{\pom}\Lb z'\Rb}{N^2_0}}\Rb\,=\,C'^2 \exp \left(-2{\sqrt{\frac{ n  }{2\, \frac{G_{\pom}\Lb z'\Rb}{N^2_0}}}}\right) \exp\Lb - \frac{1}{4\,\kappa}\ln^2\Lb 2\, \frac{G_{\pom}\Lb z'\Rb}{n\,\,N^2_0}\Rb\Rb
\eeq

For probabilities to produce $n$ gluons in the final state we have the same \eq{MD10} but with the KNO function of \eq{MDDD13}.
   ~

   ~
   
   ~
   
     \begin{boldmath}
     \section{Entropy of produced gluons }
      \end{boldmath}

      
   Since \eq{MD10} describes the  probability to find $n$-gluons in the final state the first conclusion is that the entropy is the sane in deep inelastic scattering (DIS) and for interactions of to dipoles at high energies.
   
   The    von Neumann entropy is equal 
   \beq \label{EN1}
S_E = - \sum_n \ln\Lb\mathcal{P}_n\Lb \eq{MD10} \Rb\Rb \, \mathcal{P}_n \Lb \eq{MD9} \Rb  = -\ln\Lb \frac{\bg \,z}{ e^{\bg z}}\Rb \,+\,{\rm Const} \xrightarrow{z \gg 1} \,\ln\Lb G_\pom\Lb z \Rb\Rb = \bg \,z\eeq

In other words we can state that the entropy is equal to $\ln\Lb xG(x,Q^2)\Rb$, where $xG$ is the gluon structure function.$xG$ has to be in the saturation region at high energies. This result confirms the estimates in Re.\cite{KHLE} that have been done in the simple zero dimension model for the deep inelastic scattering. On the other hand \eq{EN1} turns out 
to be much smaller than the  for QCD in  Refs.\cite{KHLE,LELAST}. 
We plan to clarify this discrepancy in our further publications.

It should be noted that the multiplicity distribution are quite different for DIS and for dipole-dipole scattering in spite of the same value of the entropy. Indeed, in the wide range of $n$ in DIS KNO function $\Psi \propto \exp\Lb - {\rm c}\, n\Rb$ while  in dipole-dipole scattering $\Psi \propto \exp\Lb -{\rm  c}\,\sqrt{n} \Rb$  in accord with Refs.\cite{MUSA,Salam}.

      ~

   ~

   ~
   
     \begin{boldmath}
     \section{Conclusions }
      \end{boldmath}

      
      In this paper we  summed the large BFKL Pomeron loops in the framework of  the BFKL Pomeron calculus. 
    Our first result is \eq{DD5} for the parton densities at high energies. This expression for $\rho_n$ stems from our attempts to reconcile the exact solution to the BK equation\cite{LETU} , which describes the rare fluctuation in the dipole-target scattering \cite{IAMU}, with the fact that this equation sums the 'fan' Pomeron diagrams of \fig{mpsi1}.  As we have mentioned  \eq{DD5} satisfies the evolution equation for the parton densities (see Ref.\cite{LELU}). Certainly,  we have to study  how these $\rho_n$ at large $Y$ will help us to solve the evolution equations and the recurrence relations for $\rho_n$ which have been discussed in Refs.\cite{LELU,LE1}.
      
  The second result is that the sum of  large Pomeron loops leads to the $S\Lb Y\Rb$ which has the same suppression  as was estimated in Ref.\cite{IAMU}.
 This result  confirms  not only the general idea of Ref.\cite{IAMU} that 'rare' fluctuation contributes to the high energy behaviour of the scattering amplitude but their size which we obtain from quite different   approach based on the BFKL Pomeron calculus.
We hope that  our result will stimulate the study of the BFKL Pomeron calculus for finding larger contribution to the scattering matrix. In particular, we have to approach the enhanced diagrams which are crucially dependant on the small Pomeron loops.

    We obtain our S-matrix from the BFKL Pomeron calculus which allow us to study the multiplicity distribution of the produced gluons  
    using AGK cutting rules.    It turns out that these secondary gluons have the  KNO  distribution. For both reaction we obtain the same average multiplicity of the produced gluons, which is equal to the gluon structure function deep in the saturation region.    
     It  leads to the entropy which  is equal to $S_E \,=\,\ln(xG(x,Q^2))$ where $xG$ is the gluon structure function. Therefore it confirms the result of Ref.\cite{KHLE}.
  
    The main theoretical problem of our approach is that the master equations (see \eq{DD2} and \eq{DD3}) are asymptotic series, which give us the analytic function. This function has a correct limit at large $z$, which describe the scattering  amplitude. However, we cannot prove that this function is a unique one. Therefore we can trust our results only at large $z$. We are planning to investigate the suggested approach in the region of $ z\,\sim\, 1$.

    We hope that this paper will contribute to the further study of the Pomeron calculus in QCD.

    ~

    ~

       {\bf Acknowledgements} 
     
   We thank our colleagues at Tel Aviv university  for
 discussions. Special thanks go A. Kovner and M. Lublinsky for stimulating and encouraging discussions on the subject of this paper. 
  This research was supported  by 
BSF grant 2022132.

~

~

~

 \appendix
\section{Saturation region: distribution of the produced gluons in the BFKL Pomeron }


In this appendix we are going to show that the produced gluons in the BFKL  Pomeron\cite{BFKL} are distributed accordingly to the Poisson law.
  Indeed, the BFKL equation has the general form in  the coordinate representation:
  \beq \label{FS1}
  \frac{\partial\,\phi_n\Lb Y,  r\Rb }{\partial\,Y}\,\,=\,\,\int d^2 r'\,K\Lb \vec{r}- \vec{r}',\vec{r}'\Rb \,\phi_{n-1}\Lb Y, r'\Rb
  \eeq
  where $\phi_n$ is the cross section of $n$ produced gluons and $K\Lb \vec{r}- \vec{r}',\vec{r}'\Rb$ is the kernel (see Ref.\cite{BFKL} for details).  In the saturation region we expect the geometric scaling behaviour for the Pomeron contribution\cite{MUT} and \eq{FS1} takes the form:
   \beq \label{FS10} 
     \frac{d\,\phi_n\Lb z\Rb }{d z}\,\,=\,\,\bg\,\phi_{n-1}\Lb Y, r'\Rb 
     \eeq  
    One can check that the solution to \eq{FS10}  can be written as   
      \beq \label{FS2}
\phi_n\Lb z\Rb\,\,=\,\,\frac{\Lb \bg\,z\Rb^n}{n!}\eeq    
  
 From \eq{FS2}   the cross section is equal to
     \beq \label{FS6}
 \sigma_{tot}\,=\,\sum_{n=0}^\infty \phi_n\Lb Y, k\Rb\,\,= \,\,e^{\bg \,z} 
   \eeq     
   while the probability to find $n$ gluons in the final state is equal to
    \beq \label{FS7}
P_n\,=\, \frac{\phi_n\Lb z\Rb}{\sigma_{tot}} \,\,=\,\, \frac{\Lb \bg\,z\Rb^n}{n!}e^{-\,\bg\,z}   \eeq    
 which is  the Poisson distribution.


\begin{thebibliography}{99} \frenchspacing
 
   \bibitem{BFKL}
   V.~S. Fadin, E.~A. Kuraev and L.~N. Lipatov,
\newblock Phys. Lett. {\bf B60}, 50 (1975);\,\,\,
E.~A. Kuraev, L.~N. Lipatov and V.~S. Fadin,
\newblock Sov. Phys. JETP {\bf 45}, 199 (1977),
\newblock [Zh. Eksp. Teor. Fiz.72,377(1977)];\,\,\,
I.~I. Balitsky and L.~N. Lipatov,
\newblock Sov. J. Nucl. Phys. {\bf 28}, 822 (1978),
\newblock [Yad. Fiz.28,1597(1978)].

 \bibitem{KLLL1}
A.~Kovner, E.~Levin, M.~Li and M.~Lublinsky,
JHEP \textbf{09} (2020), 199
[arXiv:2006.15126 [hep-ph]].
\bibitem{KLLL2}
A.~Kovner, E.~Levin, M.~Li and M.~Lublinsky,
JHEP \textbf{10} (2020), 185
[arXiv:2007.12132 [hep-ph]].


  
   \bibitem{KOLEB}
Yuri V. Kovchegov and Eugene Levin, {\it `` Quantum Chromodynamics at High Energies"}, Cambridge Monographs on Particle Physics, Nuclear Physics and Cosmology, Cambridge University Press, 2012 .
  
   
\bibitem{MUSA}
A.~H.~Mueller and G.~P.~Salam,
  Nucl.\ Phys.\ B {\bf 475}, 293 (1996),  [hep-ph/9605302];~~
  G.~P.~Salam,
  Nucl.\ Phys.\ B {\bf 461}, 512 (1996),    [hep-ph/9509353].
        \bibitem{LETU}
E.~Levin and K.~Tuchin,
  Nucl.\ Phys.\ B {\bf 573}, 833 (2000)
  [hep-ph/9908317];\,\,\,
  Nucl.\ Phys.\ A {\bf 691}, 779 (2001)
  [hep-ph/0012167]; 
   {\bf 693}, 787 (2001)
  [hep-ph/0101275].
 \bibitem{LELU}
   E.~Levin and M.~Lublinsky,
Nucl. Phys. A \textbf{730} (2004), 191-211,[hep-ph/0308279 [hep-ph];~~   
  Phys.\ Lett.\ B {\bf 607} (2005) 131, [hep-ph/0411121 [hep-ph]] ;~~ 
  Nucl.\ Phys.\ A {\bf 763} (2005) 172, [hep-ph/0501173 [hep-ph]].
   \bibitem{LIP}
 L.~N.~Lipatov,
  Sov.\ Phys.\ JETP {\bf 63}, 904 (1986)
  [Zh.\ Eksp.\ Teor.\ Fiz.\  {\bf 90}, 1536 (1986)].
\bibitem{KO1}
Y.~V.~Kovchegov,
Phys. Rev. D \textbf{72} (2005), 094009,
[arXiv:hep-ph/0508276].

     \bibitem{RS}
  P.~Rembiesa and A.~M.~Stasto,
  Nucl.\ Phys.\ B {\bf 725} (2005) 251,  [hep-ph/0503223].
  
  \bibitem{KLremark2} 
  A.~Kovner and M.~Lublinsky,
  Nucl.\ Phys.\ A {\bf 767} 171 (2006), [hep-ph/0510047].

 
  \bibitem{SHXI}  
    A.~I.~Shoshi and B.~W.~Xiao,
  Phys.\ Rev.\ D {\bf 73} (2006) 094014,  [hep-ph/0512206].
  \bibitem{KOLEV}
   M.~Kozlov and E.~Levin,
  Nucl.\ Phys.\ A {\bf 779} (2006) 142, [hep-ph/0604039].
\bibitem{nestor} N. Armesto, S. Bondarenko, J. G. Milhano and P. Quiroga, JHEP 0805 (2008) 103, arXiv:0803.0820 [hep-ph]. 
  
\bibitem{LEPRI}
  E.~Levin and A.~Prygarin,
  Eur.\ Phys.\ J.\ C {\bf 53} (2008) 385,  [hep-ph/0701178].
  
  
  
  
\bibitem{LMM}
A.~D.~Le, A.~H.~Mueller and S.~Munier,
Phys. Rev. D \textbf{104}, 034026 (2021),
[arXiv:2103.10088 [hep-ph]].
\bibitem{LEM}
E.~Levin,
Phys. Rev. D \textbf{104}, no.5, 056025 (2021),
[arXiv:2106.06967 [hep-ph]].

 
     
  
  
     \bibitem{MUT}
A.~H.~Mueller and D.~N.~Triantafyllopoulos,
  Nucl.\ Phys.\ B {\bf 640} (2002) 331,
  [hep-ph/0205167]
  \bibitem{MUPE}
S.~Munier and R.~B.~Peschanski,
Phys. Rev. D \textbf{69} (2004), 034008
doi:10.1103/PhysRevD.69.034008
[arXiv:hep-ph/0310357 [hep-ph]]; 
Phys. Rev. Lett. \textbf{91} (2003), 232001
doi:10.1103/PhysRevLett.91.232001
[arXiv:hep-ph/0309177 [hep-ph]].
  
  
      \bibitem{IIML}
   E.~Iancu, K.~Itakura and L.~McLerran,
  Nucl.\ Phys.\ A {\bf 708} (2002) 327,
  [hep-ph/0203137]. 
  
 \bibitem{LIREV}
                L.~N.~Lipatov,
  Phys.\ Rept.\  {\bf 286} (1997) 131, [hep-ph/9610276].
               
               
               
\bibitem{LIFT}
  L.~N.~Lipatov,
  Nucl.\ Phys.\ B {\bf 365}, 614 (1991),
  Nucl.\ Phys.\ B {\bf 452}, 369 (1995), [arXiv:hep-ph/9502308];\\
R.~Kirschner, L.~N.~Lipatov and L.~Szymanowski,
  Nucl.\ Phys.\ B {\bf 425}, 579 (1994), [arXiv:hep-th/9402010];
  Phys.\ Rev.\ D {\bf 51}, 838 (1995), [arXiv:hep-th/9403082].

    
          
                
\bibitem{GLR} 
L.~V.~Gribov, E.~M.~Levin and M.~G.~Ryskin,
  Phys.\ Rept.\  {\bf 100}, 1 (1983).
 \bibitem{GLR1}
E.~M.~Levin and M.~G.~Ryskin,
  Phys.\ Rept.\  {\bf 189}, 267 (1990).



\bibitem{MUQI}
A. H. Mueller and J. Qiu, 
Nucl. Phys. {\bf B268} (1986) 427.
 
\bibitem{MUDI}
  A.~H.~Mueller,
  Nucl.\ Phys.\ B {\bf 415} (1994) 373;\,\,\,
  Nucl.\ Phys.\ B {\bf 437} (1995) 107,  [hep-ph/9408245]\\
      A.~H.~Mueller and B.~Patel, 
      Nucl. Phys. B 425, 471, 1994.  

\bibitem{Salam}
  G.~P.~Salam,
  Nucl.\ Phys.\ B {\bf 461}, 512 (1996);
   [hep-ph/9509353]. 
             
\bibitem{NAPE}
H.~Navelet and R.~B.~Peschanski,
Nucl. Phys. B \textbf{507} (1997), 353-366
[arXiv:hep-ph/9703238].

               
   \bibitem{BART}
J.~Bartels,
Z. Phys. C \textbf{60} (1993), 471-488
;~~
J.~Bartels and M.~Wusthoff,
Z. Phys. C \textbf{66} (1995), 157-1801;
~~ J.~Bartels and C.~Ewerz,
JHEP \textbf{09} (1999), 026
[arXiv:hep-ph/9908454];~~
   C.~Ewerz,
  JHEP {\bf 0104} (2001) 031,  [hep-ph/0103260].\,\,\,
  
\bibitem{BKP}
J.~Bartels,
%
Nucl.\ Phys.\  {\bf B175}, 365 (1980);\\
J.~Kwiecinski and M.~Praszalowicz,
%
Phys.\ Lett.\  {\bf B94}, 413 (1980).

    \bibitem{MV}
L. McLerran and R. Venugopalan, 
Phys. Rev. {\bf D49} (1994) 2233,
Phys. Rev. {\bf D49} (1994), 3352;   
{\bf D50} (1994) 2225;
 {\bf D59} (1999) 09400.  
   
  
   \bibitem{KOLE}
  Y.~V.~Kovchegov and E.~Levin,
  Nucl.\ Phys.\ B {\bf 577} (2000) 221,  [hep-ph/9911523].
  
   \bibitem{BRN}
M. A. Braun,
Eur. Phys. J. {\bf C16} (2000) 337, [arXiv:hep-ph/0001268]; \\
M. A. Braun and G. P. Vacca,
Eur. Phys. J. {\bf C6} (1999) 147,[arXiv:hep-ph/9711486];\\
J.~Bartels, M.~Braun and G.~P.~Vacca,
  Eur.\ Phys.\ J.\ C {\bf 40}, 419 (2005), [arXiv:hep-ph/0412218],
\\
 J.~Bartels, L.~N.~Lipatov and G.~P.~Vacca,
  Nucl.\ Phys.\ B {\bf 706}, 391 (2005),  [arXiv:hep-ph/0404110].

\bibitem{BRAUN}  M.~A.~Braun,
  Phys.\ Lett.\ B {\bf 483}, 115 (2000), [hep-ph/0003004];
  Eur.\ Phys.\ J.\ C {\bf 33}, 113 (2004),[hep-ph/0309293];
  Phys.\ Lett.\ B {\bf 632}, 297 (2006). 
  
  
  
   
  \bibitem{B}
I.~Balitsky,
{Phys.\ Rev.} {\bf D60}, 014020 (1999),[arXiv:hep-ph/9812311];\,\,\,\,
\bibitem{K}
Y.~V.~Kovchegov,
{Phys.\ Rev.}  {\bf D60}, 034008  (1999),[arXiv:hep-ph/9901281].
   \bibitem{KOLU}
A.~Kovner and M.~Lublinsky,
JHEP \textbf{02} (2007), 058,
[arXiv:hep-ph/0512316 [hep-ph]].
\bibitem{JIMWLK1}
J.~Jalilian-Marian, A.~Kovner, A.~Leonidov and H.~Weigert
  \href{http://dx.doi.org/10.1016/S0550-3213(97)00440-9}, Nucl. Phys. {\bf
  B504} (1997)  415--431,
\href{http://arxiv.org/abs/hep-ph/9701284}[ arXiv:hep-ph/9701284].

\bibitem{JIMWLK2}
J.~Jalilian-Marian, A.~Kovner, A.~Leonidov and H.~Weigert 
  \href{http://dx.doi.org/10.1103/PhysRevD.59.014014}, Phys.Rev. {\bf D59}
  (1998)  014014,
\href{http://arxiv.org/abs/hep-ph/9706377}[arXiv:hep-ph/9706377
  [hep-ph]].

\bibitem{JIMWLK3}
A.~Kovner, J.~G. Milhano and H.~Weigert
  \href{http://dx.doi.org/10.1103/PhysRevD.62.114005}, Phys. Rev. {\bf
  D62} (2000)  114005,
\href{http://arxiv.org/abs/hep-ph/0004014}[ arXiv:hep-ph/0004014].

\bibitem{JIMWLK4}
E.~Iancu, A.~Leonidov and L.~D. McLerran
  \href{http://dx.doi.org/10.1016/S0375-9474(01)00642-X},Nucl. Phys. {\bf
  A692} (2001)  583--645,
\href{http://arxiv.org/abs/hep-ph/0011241}[ arXiv:hep-ph/0011241].

\bibitem{JIMWLK5}
E.~Iancu, A.~Leonidov and L.~D. McLerran 
  \href{http://dx.doi.org/10.1016/S0370-2693(01)00524-X}, Phys. Lett. {\bf
  B510} (2001)  133--144,
\href{http://arxiv.org/abs/hep-ph/0102009}[ arXiv:hep-ph/0102009].

\bibitem{JIMWLK6}
E.~Ferreiro, E.~Iancu, A.~Leonidov and L.~McLerran
  \href{http://dx.doi.org/10.1016/S0375-9474(01)01329-X}, Nucl. Phys. {\bf
  A703} (2002)  489--538,
\href{http://arxiv.org/abs/hep-ph/0109115}[ arXiv:hep-ph/0109115].



\bibitem{JIMWLK7}
 H.~Weigert,
Nucl. Phys. A \textbf{703} (2002), 823-860
[arXiv:hep-ph/0004044 [hep-ph]].
\bibitem{JIMWLK8}
 A.~Kovner and J.~G.~Milhano,
Phys. Rev. D \textbf{61} (2000), 014012
[arXiv:hep-ph/9904420 [hep-ph]].

\bibitem{AKLL}
T.~Altinoluk, A.~Kovner, E.~Levin and M.~Lublinsky,
JHEP \textbf{04} (2014), 075
[arXiv:1401.7431 [hep-ph]].
\bibitem{KOLU11}
A.~Kovner and M.~Lublinsky,
Phys. Rev. D \textbf{71} (2005), 085004
[arXiv:hep-ph/0501198 [hep-ph]].
\bibitem{KOLUD} 
  A.~Kovner and M.~Lublinsky,
  Phys.\ Rev.\ Lett.\  {\bf 94}, 181603 (2005), [hep-ph/0502119].


\bibitem{BA05}
  I.~Balitsky,
   Phys.\ Rev.\ D {\bf 72}, 074027 (2005), arXiv:hep-ph/0507237.

\bibitem{SMITH}
  Y.~Hatta, E.~Iancu, L.~McLerran, A.~Stasto and D.~N.~Triantafyllopoulos,
  Nucl.\ Phys.\ A {\bf 764}, 423 (2006),[arXiv:hep-ph/0504182].
  \bibitem{KLW}
A.~Kovner, M.~Lublinsky and U.~Wiedemann,
JHEP \textbf{06} (2007), 075
[arXiv:0705.1713 [hep-ph]];
~~~T.~Altinoluk, A.~Kovner, M.~Lublinsky and J.~Peressutti,
  JHEP {\bf 0903}, 109 (2009),  [arXiv:0901.2559 [hep-ph]].





\bibitem{kl} A. Kovner and M. Lublinsky
Nucl. Phys. A \textbf{767}, 171-188 (2006),[arXiv:hep-ph/0510047].


 


\bibitem{LEPR}
A.~Kormilitzin, E.~Levin and A.~Prygarin,
Nucl. Phys. A \textbf{813} (2008), 1-13
[arXiv:0807.3413 [hep-ph]];\,\,\,
E.~Levin and A.~Prygarin,
Eur. Phys. J. C \textbf{53} (2008), 385-399
[arXiv:hep-ph/0701178 [hep-ph]].

 \bibitem{LE1}
E.~Levin,
Phys. Rev. D \textbf{107} (2023) no.5, 054012
[arXiv:2209.07095 [hep-ph]].
\bibitem{LE2}
E.~Levin,
``Scattering amplitude in QCD: summing large Pomeron loops,''
[arXiv:2403.10364 [hep-ph]].
\bibitem{LE11}
E.~Levin,
Nucl. Phys. A \textbf{763} (2005), 140-171,
[arXiv:hep-ph/0502243 [hep-ph]].
  \bibitem{IAMU}
E.~Iancu and A.~Mueller,
Nucl. Phys. A \textbf{730} (2004), 494-513
[arXiv:hep-ph/0309276 [hep-ph]];

 
 \bibitem{IAMU1}
E.~Iancu and A.~Mueller\,\,
Nucl. Phys. A \textbf{730} (2004), 460-493
[arXiv:hep-ph/0308315].
\bibitem{GS}
J.~Bartels, E.~Levin,
  Nucl.\ Phys.\  {\bf B387 } (1992)  617-637;\,\,
 A.~M.~Stasto, K.~J.~Golec-Biernat, J.~Kwiecinski,
  Phys.\ Rev.\ Lett.\  {\bf 86 } (2001)  596-599,
  [hep-ph/0007192];\,\,\,L.~McLerran, M.~Praszalowicz,
  Acta Phys.\ Polon.\  {\bf B42 } (2011)  99,
  [arXiv:1011.3403 [hep-ph]]  {\bf B41 } (2010)  1917-1926,
  [arXiv:1006.4293 [hep-ph]].
  


 \bibitem{AGK}
V.~A.~Abramovsky, V.~N.~Gribov and O.~V.~Kancheli,
Yad. Fiz. \textbf{18} (1973), 595, (Sov.J. Nucl.Phys. 18 (1974),308);
\bibitem{KNO} A.M. Polyakov, 
        Zh. Eksp. Teor. Fiz. {\bf 59}, 542 (1970);~~
 Z. Koba, H.B. Nielsen and P. Olesen, 
	Nucl. Phys. {\bf B40}, 317 (1972); 
~~~ Z. Koba,
	in Proc. of the 1973 CERN School of Physics,
	p.~171, CERN Yellow Report CERN-73-12 (1973)

\bibitem{KHLE}
D.~E.~Kharzeev and E.~M.~Levin,
Phys. Rev. D \textbf{95} (2017) no.11, 114008,
[arXiv:1702.03489 [hep-ph]].



\bibitem{BOREL}
 	Bruce Shawyer and Bruce Watson, {\it ``Borel's Method of Summability, Theory and Application''} , Clarendon Press, Oxford,1994;\\
	Ovidiu Costin,{``Asymptotocs and Borel Summability''}, Chapman \& HALL/CRC Monographs and Surveys in Pure and Applied Mathematics, CRC Press, Taylor \& Frencis Group,2009;\\$http://www1.phys.vt.edu/~ersharpe/spec-fn/app-d.pdf$.
\bibitem{CLMSOFT}
C.~Contreras, E.~Levin and M.~Sanhueza,
Phys. Rev. D \textbf{104} (2021) no.11, 116020,
[arXiv:2106.06214 [hep-ph]].
  \bibitem{LEHP}
E.~Levin,
Phys. Rev. D \textbf{49} (1994), 4469-4480.
\bibitem{LELAST}
E.~Levin,
Eur. Phys. J. C \textbf{84} (2024) no.7, 662
[arXiv:2306.12055 [hep-ph]].



\end{thebibliography}
\end{document}